\documentclass[preprints,instruments,review,submit,oneeauthor,pdftex]{Definitions/mdpi} 
\usepackage{mathrsfs} 
\firstpage{1} 
\pubvolume{xx}
\issuenum{1}
\articlenumber{5}
\pubyear{2019}
\copyrightyear{2019}
\history{Received: date; Accepted: date; Published: date}

\pdfoutput=1

\Title{Calibration of calorimetric measurement in a liquid argon time projection chamber}

\Author{Tingjun Yang $^{1}$\orcidA{}}

\AuthorNames{Tingjun Yang}

\address{%
$^{1}$ \quad Fermi National Accelerator Laboratory, Batavia, IL 60510, USA; tjyang@fnal.gov
}

\abstract{The liquid argon time projection chamber provides high resolution event images and excellent calorimetric resolution for studying neutrino physics and searching for beyond standard model physics. In this article, we review the main physics processes that affect the detector response, including the electronics and field responses, space charge effects, electron attachment to impurities, diffusion and recombination. We describe methods to measure those effects, which are used to calibrate the detector response and convert the measured raw ADC counts to the original energy deposition. 
}


\begin{document}
\tableofcontents


\section{Introduction}
The liquid argon time projection chamber (LArTPC) detector technology provides high resolution event images and excellent calorimetric resolution for particle identification. The charged particles produce ionization electrons and scintillation light when they traverse liquid argon. The ionization electrons are drifted towards the anode planes under the electric field. In the single-phase design, the moving electrons produce current on the TPC wires at the anode. The wire signal is amplified and digitized by the electronics and then read out by the data acquisition system. The goal of the detector calibration is to convert the raw signal to the original energy deposition by removing the detector and physics effects, which we will discuss in detail in this paper. The calorimetric information is crucial for particle identification in a LArTPC, such as separating minimum ionizing particles (muons and charged pions) from highly ionizing particles (kaons and protons) and separating electrons from photons, which is the base for the neutrino cross section and oscillation measurements and the searches for beyond standard model physics. This paper summarizes the common techniques used to calibrate the TPC signal. The calibration of photon detector signal is not discussed in this paper. 
 
\section{TPC Signal formation}
\label{sec:sigform}
The electron-ion pairs ($e^{-}$, Ar$^{+}$) are produced from energy loss by charged particles in liquid argon through the ionization process:
\begin{equation}
  N_{i} = \frac{\Delta E}{W_{\textrm{ion}}},
\end{equation}
where $N_{i}$ is the number of electron-ion pairs, $\Delta E$ is the energy loss, and $W_{\textrm{ion}} = 23.6 \pm 0.3$ eV~\cite{Miyajima:1974zz} is the ionization work function.

Some of the ionization electrons are recombined with surrounding molecular argon ions to form the excimer Ar$^{*}_{2}$. The excimers from both argon excitation and electron-ion recombination undergo dissociative decay to their ground state by emitting a vacuum ultraviolet photon. The free electrons that escape electron-ion recombination are drifted towards the wire planes under the electric field. Several effects can affect the electron drift. The electrons can be attached to contaminants in the liquid argon such as oxygen and water, which causes an attenuation of the signal on the TPC wires. The electron cloud is smeared both in the longitudinal and transverse directions by the diffusion effect. For a LArTPC located on the surface, there is a large flux of cosmic ray muons in the detector volume. Because of this, there is a significant accumulation of slow-moving argon ions (space charge) inside the detector, which leads to a position-dependent distortion to the electric field. The distorted electric field changes the electron-ion recombination rate and the trajectory of the free drifting electrons.

In a LArTPC with one grid plane and three instrumented wire planes, the drifting electrons first pass the grid plane, then pass two induction planes and eventually are collected by the wires on the collection plane. The electrons produce bipolar signals on the induction wires and unipolar signals on the collection wires. The wire signals is amplified by a preamplifier and then digitized by an analog-to-digital converter (ADC). 
\section{LArTPC calibration}

In order to measure the energy loss per unit length ($dE/dx$) for particle identification, we need to convert the measured ADC counts to the energy deposition. The detector calibration procedure needs to remove all the instrument and physics effects following the reverse order of the signal formation described in Sec.~\ref{sec:sigform}. We discuss different ways to  measure each effect and practical methods to remove them.  

\subsection{Electronics response and field response}
\label{sec:electronics}
\begin{figure}[htp!]
  \centering
  \includegraphics[width=0.70\textwidth]{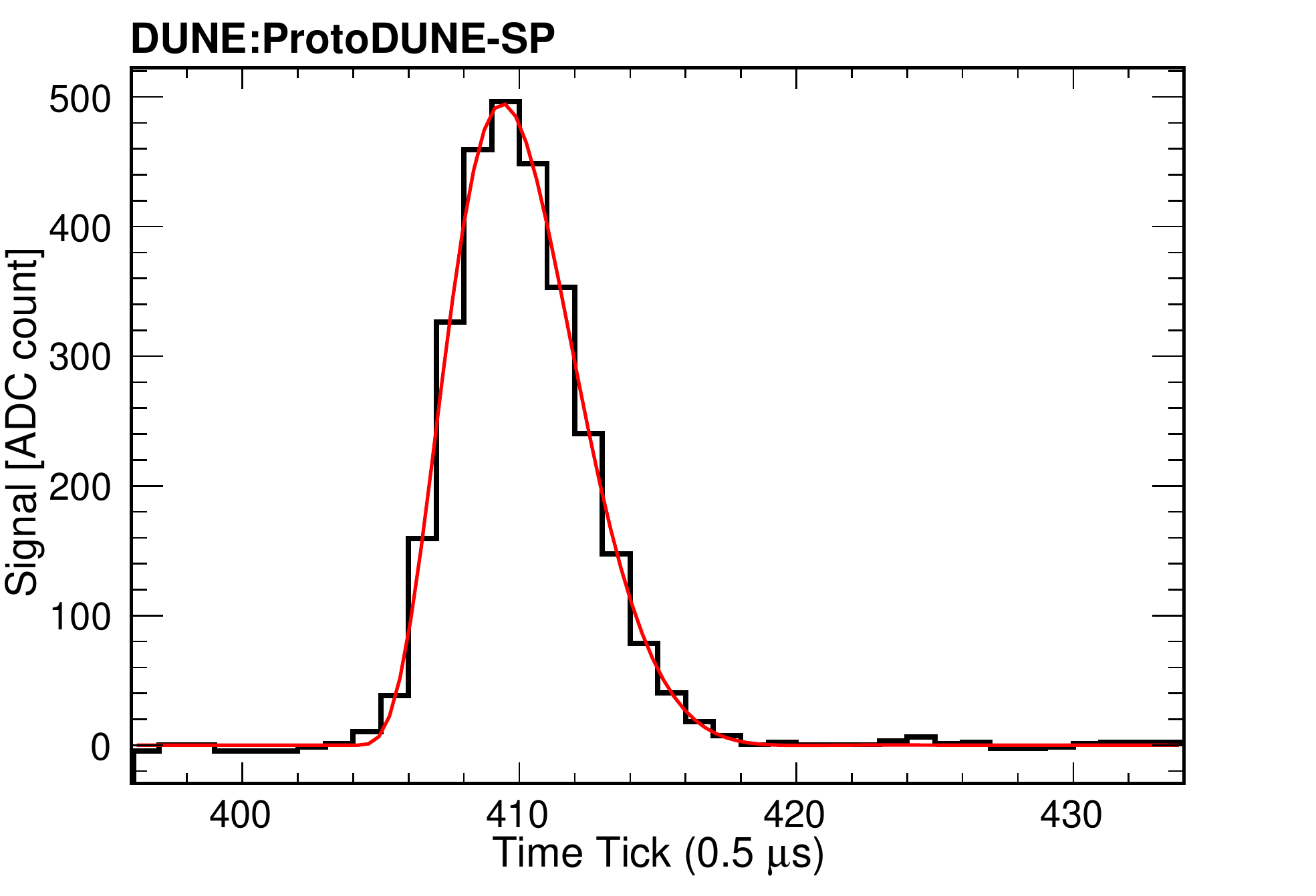}
\caption{
Response of the ProtoDUNE electronics to an input charge of DAC setting 3. The red curve is a fit to a simulated function of electronics response. Plot courtesy of David Adams (BNL) and the DUNE collaboration.
}
\label{fig:pulser}
\end{figure}

Many LArTPC experiments have an electronics calibration system that has the capability to inject a known charge into each of the amplifiers connected to the TPC wires. In the ProtoDUNE pulser calibration system~\cite{Abi:2020mwi}, the injected charge is controlled by a six-bit voltage digital-to-analog converter (DAC):
\begin{equation}
  Q = SQ_{s},
\end{equation}
where $S$ is the DAC setting (0 - 63) and the step charge $Q_{s}$ = 3.43~fC~$\sim$ 21,400 electrons, which is roughly equivalent to the charge deposition of a minimum ionizing particle traveling parallel to the wire plane and perpendicular to that plane's wire direction on a single wire. The charge calibration is expressed as a gain for each channel normalized such that the product of the gain and the integral of the ADC counts over the pulse (pulse area) in a collection channel gives the charge in the pulse, i.e.~$Q = g A$. Figure~\ref{fig:pulser} shows the response of the ProtoDUNE electronics to an input charge of DAC setting 3. 

\begin{figure}[htp!]
  \centering
  \includegraphics[width=0.70\textwidth]{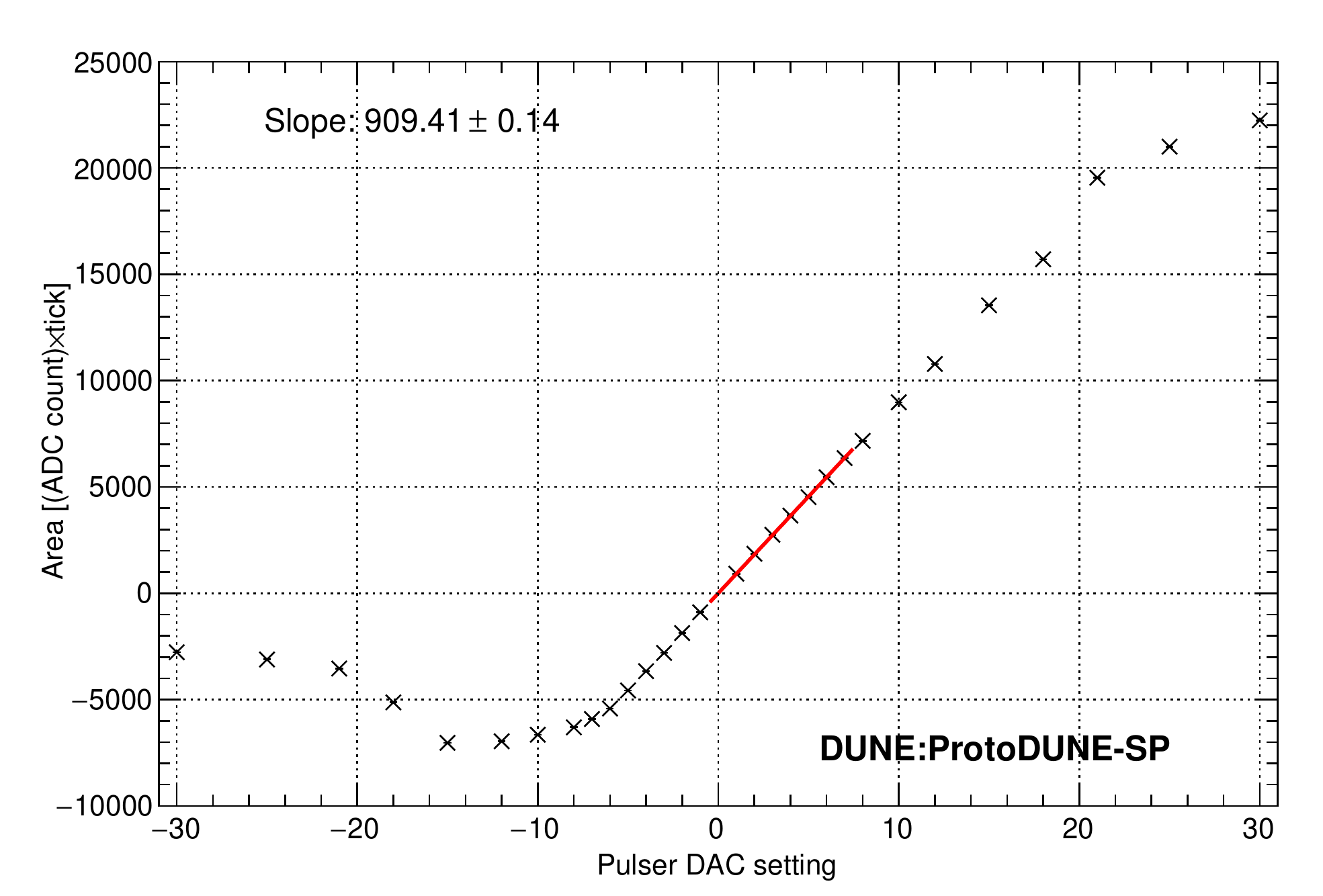}
\caption{
Measured pulse area vs.\ DAC setting for a typical collection channel.
The red line shows the fit used to extract the gain. This figure is from~\cite{Abi:2020mwi}.
}
\label{fig:tpcarea}
\end{figure}

Figure~\ref{fig:tpcarea} shows the measured pulse area vs. DAC setting and
the fit for a typical collection channel.
The response is fairly linear over the DAC setting range (-5,~20) with saturation setting in
outside this range. Typical track charge deposits are one to four times the step charge and this saturation is only an issue for very heavily ionizing tracks. The gain for this channel is $g = (21.4$~ke)/(909.4~(ADC~count)-tick) =~23.5~e/((ADC~count)-tick).

\begin{figure}[htp!]
\centering
  \includegraphics[width=0.70\textwidth]{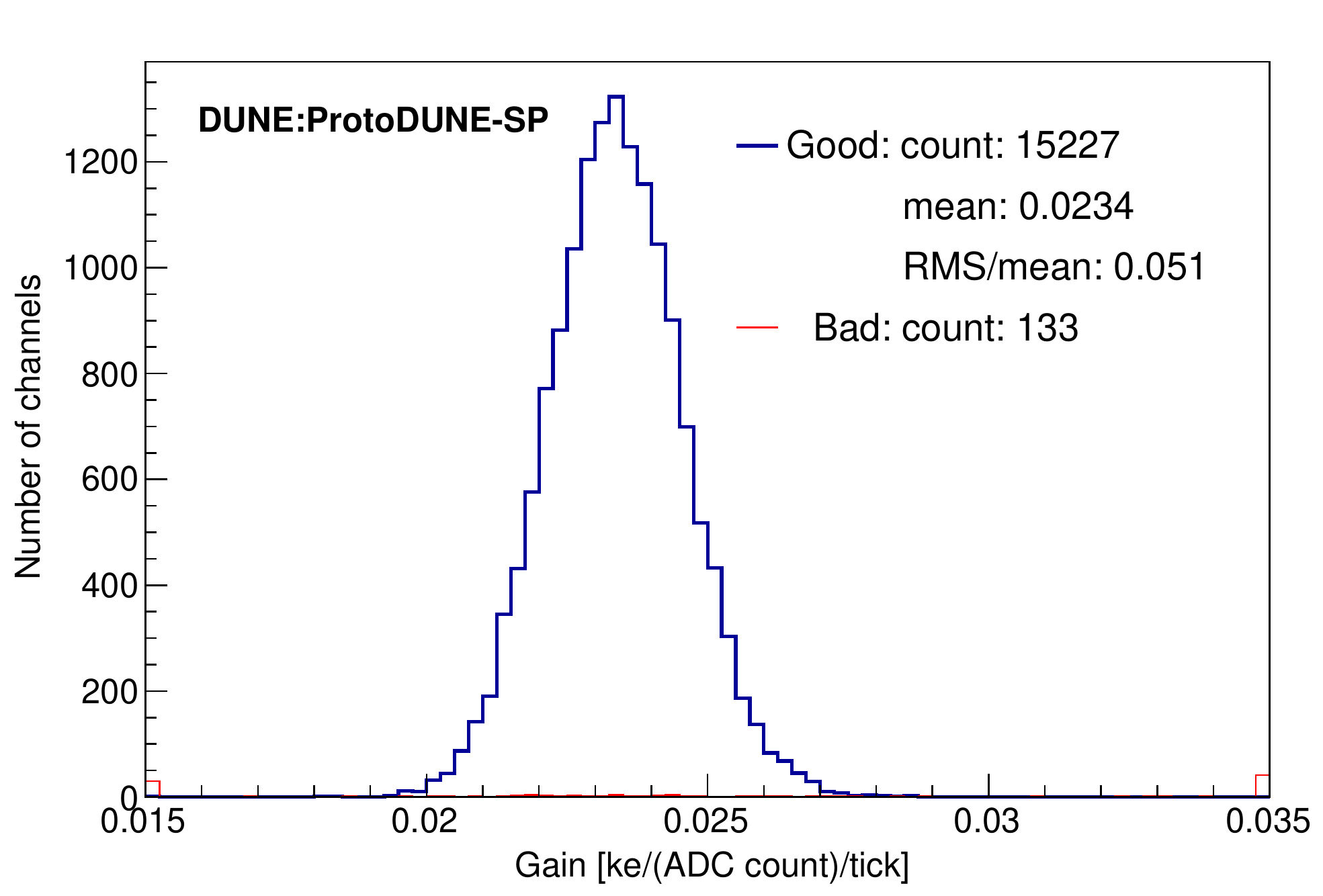}
\caption{
Distribution of fitted gains for good (blue) and bad/noisy (red) channels.
The legend indicates the number of channels in each category and gives the mean (23.4~e/(ADC~count)/tick)) and RMS/mean (5.1\%) for the good channels. This figure is from~\cite{Abi:2020mwi}.
}
\label{fig:tpcgains}
\end{figure}

Figure~\ref{fig:tpcgains} shows the distribution of these gains for all channels.
Channels flagged as bad or especially noisy in an independent hand scan are shown
separately.
The gains for the remaining good channels are contained in a narrow peak with an
RMS of 5.1\% reflecting channel-to-channel response variation in the ADCs
and gain and shaping time variations in the amplifiers. 

\begin{figure}[htp!]
\centering
  \includegraphics[width=0.70\textwidth, trim={0cm 8cm 3cm 4.5cm},clip]{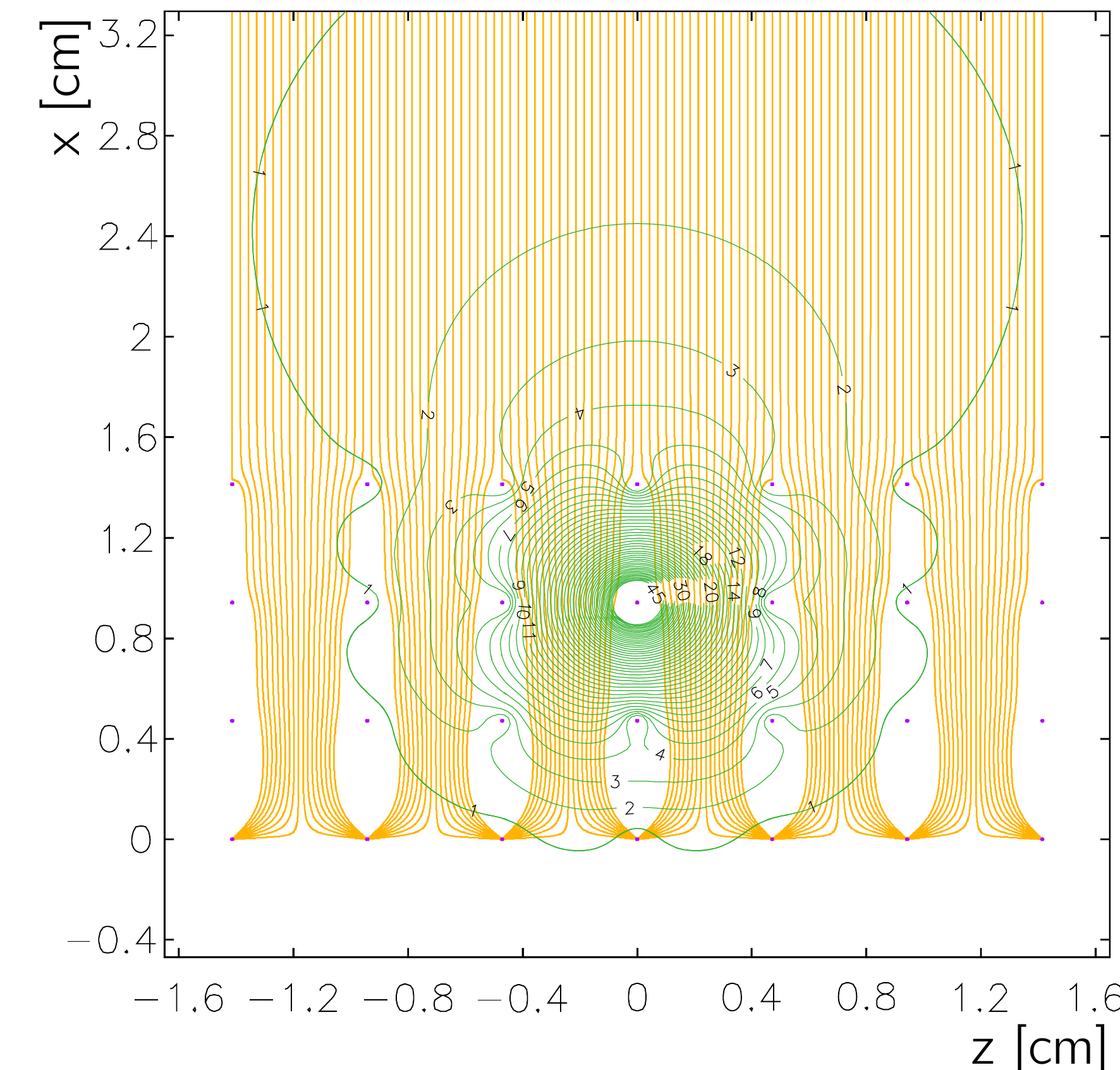}
\caption{
Garfield simulation of electron drift paths (yellow lines) in a 2D ProtoDUNE-SP TPC scheme and the equal weighting potential lines (green) for a given wire in the first induction plane, where the latter is shown in percentage from 1\% to 45\%. This figure is from~\cite{Abi:2020mwi}
}
\label{fig:garfield}
\end{figure}

Ionization electrons follow the electric field lines as they pass through the wire planes producing direct and induced signals on nearby wires. Based on the Shockley-Ramo theorem, the instantaneous electric current $i$ on a particular electrode (wire) which is held at constant voltage, is given by
\begin{equation}
  i = e\nabla\phi \cdot \vec{v}_e,
\end{equation}
where $e$ is the charge in motion, and $\vec{v}_e$ is the
charge velocity at a given location, which is determined by the wire bias voltage settings. The weighting potential $\phi$
of a selected electrode at a given location is determined by virtually
removing the charge and setting the potential of the selected
electrode to unity while grounding all other conductors. For ProtoDUNE, the drift electric field and the weighting potential are calculated with Garfield~\cite{garfield}, as shown in Figure~\ref{fig:garfield}. The weighting field determines the bipolar signal shape on the induction plane and the unipolar signal shape on the collection plane.

The electronics response and the field response are normally removed from the raw wire signal through deconvolution~\cite{Baller:2017ugz}. If we ignore the charge induction on the neighboring wires, the time sampled wire signal read out by the data acquisition system, denoted $W(t)$, is the convolution of the ionization charge approaching the wire plane, $Q(t)$, with the field response $F(t)$, and the electronics response $E(t)$, plus the noise $N(t)$:
\begin{equation}
  W(t) = Q(t)*F(t)*E(t) + N(t).
\end{equation}
The time-dependent ionization charge can be recovered by deconvolution:
\begin{equation}
  Q^\prime(t) = \mathcal F^{-1}  \Bigg[ \frac{\Phi(f) \cdot {\mathcal F}(W(t))}{ {\mathcal F}(F(t)) \cdot {\mathcal F}(E(t))} \Bigg]
\end{equation}
where $\mathcal F$ is the Fourier Transform and $\Phi (f)$ is a filter function that maximizes the signal to noise ratio. The integral of filter function is usually normalized to 1 so it does not affect the charge reconstruction. The integral of electronics response is set to the gain measured by the pulser calibration system. For the collection wires, the integral of field response is normalized to 1. Therefore, the deconvolved signal $Q(t)$ is the ionization charge at the wire. For the induction wires, the reconstruction of charge is more difficult because of the bipolar signal shape. 

A more advanced signal reconstruction technique is a two-dimensional (2D) deconvolution involving both the time and wire dimensions. The 2D deconvolution takes into account charge induction on the neighboring wires, which gives more accurate ionization charge information. It was first developed for the MicroBooNE experiment~\cite{Adams:2018dra} and then successfully used by the ProtoDUNE experiment~\cite{Abi:2020mwi}.

\subsection{Space Charge Effects}
\label{sec:sce}
In a large LArTPC located on the surface, space charge effects (SCE), i.e. the accumulation of positive argon ions ($Ar_{2}^{+}$) produced by the cosmic rays, distort the electric field significantly. Due to the ion mobility ($\mu_{i}\sim$ 10$^{-3} $cm$^{2}$V$^{-1}s^{-1}$) much smaller than the free electron one ($\mu_{e} \sim$ 500 cm$^{2}$V$^{-1}$s$^{-1}$), positive ions can survive in the drift region of the TPC for several minutes before being neutralized on the cathode or on the field shaping electrodes. In Ref.~\cite{Palestini:1998an}, the authors calculated the electric field $\mathscr{E}$ and space charge $\rho^{+}$, which we briefly summarize here. In the simplified model, $\mathscr{E}$ and $\rho^{+}$ are determined by the Maxwell and charge continuity equations:
\begin{equation}
  \frac{\partial\mathscr{E}}{\partial x} = \frac{\rho^{+}}{\varepsilon_{r}\varepsilon_{0}},
  \label{eq:maxwell}
\end{equation}
\begin{equation}
  \frac{\partial\rho^{+}}{\partial t} + \frac{\partial(\rho^{+} v_{i})}{\partial x} = J,
  \label{eq:continuity}
\end{equation}
where $x$ is the drift coordinate normal to the wire planes ($x = 0$ and $x = D$ define the anode and the cathode positions, respectively, and $D$ is the distance between anode and cathode), $\varepsilon_{0} = 8.854$ pF/m is the permittivity of vacuum, $\varepsilon_{r} = 1.504$ is the relative permittivity (dielectric constant) of liquid argon. The parameter $J$ is the average injected charge, which depends on the cosmic ray rate and the electric field applied. The ICARUS experiment measured $J = (1.9\pm0.1)\times10^{-10}$ C m$^{-3}$ s$^{-1}$ at the nominal electric field of 500 V/cm~\cite{Antonello:2020qht}. The MicroBooNE experiment estimated $J = 1.6\times10^{-10}$ C m$^{-3}$ s$^{-1}$ at the lower nominal electric field of 273 V/cm~\cite{Abratenko:2020bbx}. The parameter $v_{i} = \mu_{i}\mathscr{E}$ is the ion drift velocity. There is an uncertainty in the value of the mobility of positive argon ions $\mu_{i}$ (see references in~\cite{Palestini:2020dhv}). The measured mobility depends on several experimental conditions, such as the impurity level and temperature. A recent ICARUS measurement suggested $\mu_{i}$ is consistent with $0.9\times10^{-3}$ cm$^{2}$V$^{-1}$s$^{-1}$. MicroBooNE showed a reasonable agreement between the model prediction and the measured electric field~\cite{Abratenko:2020bbx} if a $\mu_{i}$ of $1.5\times10^{-3}$ cm$^{2}$V$^{-1}$s$^{-1}$ as reported in~\cite{doi:10.1063/1.334552} is used.

Eq.~\ref{eq:maxwell} and eq.~\ref{eq:continuity} are a one-dimensional approximation, which assumes symmetry in y and z coordinates. Introducing the dimensionless variable $\alpha$:
\begin{equation}
  \alpha = \frac{D}{\mathscr{E}_{0}}\sqrt{\frac{J}{\varepsilon_{r}\varepsilon_{0}\mu_{i}}},
\end{equation}
where $\mathscr{E}_{0} = V/D$ is the nominal electric field in absence of space charge produced by the high voltage $V$, eq.~\ref{eq:maxwell} and eq.~\ref{eq:continuity} can be solved to give:
\begin{equation}
  \mathscr{E}(x) = \mathscr{E}_{0}\sqrt{\left(\frac{\mathscr{E}_{a}}{\mathscr{E}_{0}}\right)^2+\alpha^{2}\frac{x^{2}}{D^{2}}},
\end{equation}
\begin{equation}
  \rho^{+}(x) = \frac{Jx}{\mu_{i}\mathscr{E}(x)},
\end{equation}
where $\mathscr{E}_a$ denotes the field at anode and is determined by the integral:
\begin{equation}
  \int_{0}^{D} \mathscr{E}(x) \,dx = V.
\end{equation}

\begin{figure}[htp!]
\centering
  \includegraphics[width=0.70\textwidth]{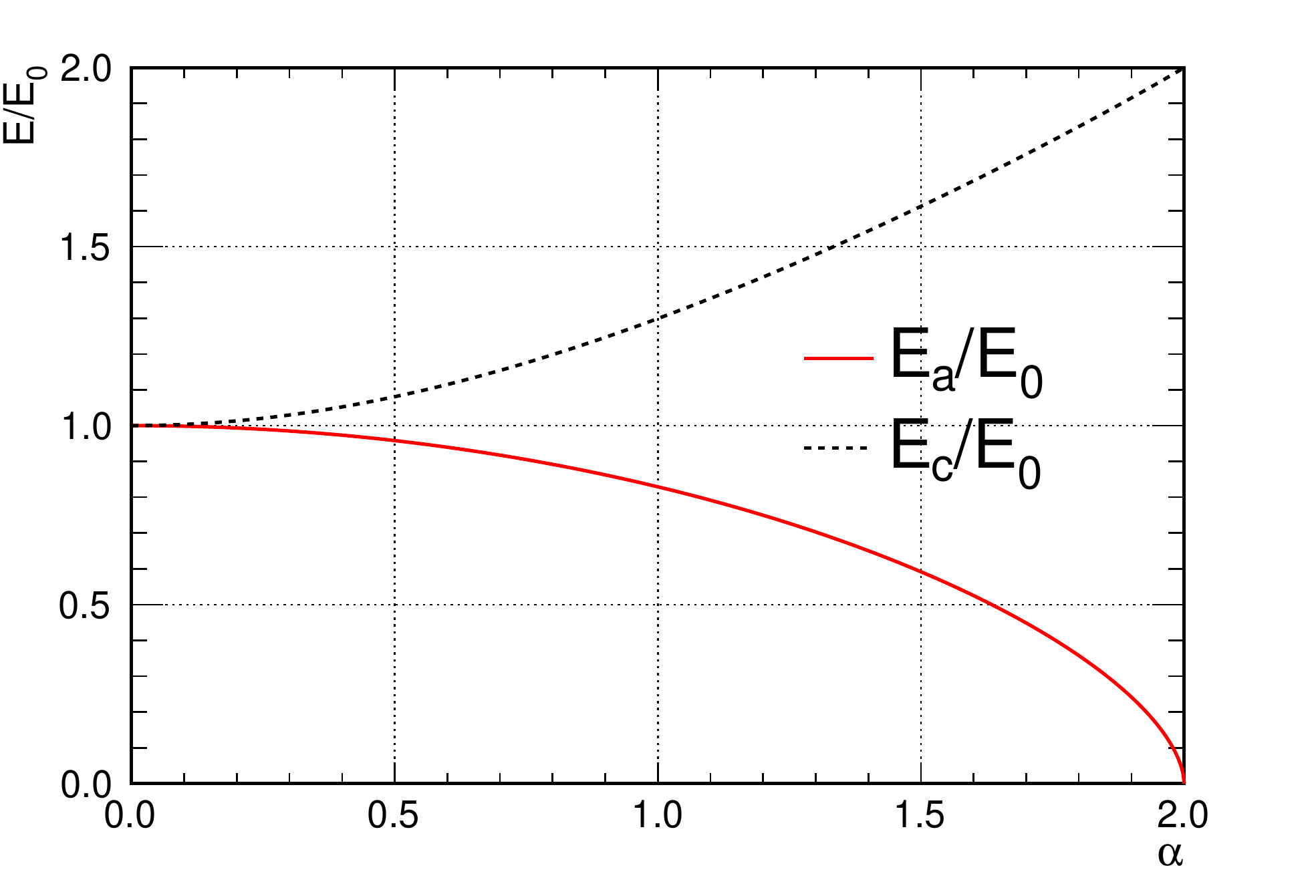}
\caption{
Normalized electric field at the anode $\mathscr{E}_{a}/\mathscr{E}_{0}$ and cathode $\mathscr{E}_{c}/\mathscr{E}_{0}$ as a function of the parameter $\alpha$, reproduced from Ref.~\cite{Palestini:1998an}.
}
\label{fig:Er}
\end{figure}

The space charge increases the electric field at the cathode ($\mathscr{E}_{a}$) and decreases the electric field at the anode ($\mathscr{E}_{c}$) as shown in Figure~\ref{fig:Er}. When $\alpha\geq2$, the electric field at anode drops to 0 and the TPC stops working.

There are many other effects that could affect the electric field. If the liquid argon purity is low, the contribution of negative ions from electrons attaching to the electronegative contaminants such as oxygen and water is not negligible~\cite{Luo:2020itx}. The fluid flow caused by the recirculation of liquid argon through the filter materials changes the space charge distribution since the fluid flow velocity and the ion drift velocity are of the same order. More sophisticated calculation of the space charge effects is discussed in~\cite{Palestini:1998an, Palestini:2020dhv}.

We now compare the electric field distortion in the three LArTPCs located on the surface: ICARUS, MicroBooNE and ProtoDUNE. Table~\ref{tab:sce} shows the running conditions of the three experiments. The calculated electric field distortion is systematically higher than the measurements. The fact that MicroBooNE prefers a larger value of positive argon ion mobility $\mu_{i}$ could be due to the different running conditions (e.g. argon purity and electric field). A careful treatment of detector-specific SCE is important for calibration.
\begin{table}[htp!]
\caption{Running conditions of the ICARUS, MicroBooNE and ProtoDUNE experiments and calculated electric field distortion in comparison with the measurements.}
\centering
\begin{tabular}{c|ccc}
\toprule
& \textbf{ICARUS}	& \textbf{MicroBooNE} & \textbf{ProtoDUNE}\\
\midrule
HV (kV) & 75 & 70 & 180 \\
$D$ (m) & 1.5 & 2.56 & 3.6\\
$\mathscr{E}_{0}$ (V/cm) & 500 & 273 & 500 \\
$J$ ($10^{-10}$ C m$^{-3}$ s$^{-1}$) & 1.9 & 1.6 & 1.9 \\
$\mu_{i}$ ($10^{-3}$ cm$^{2}$V$^{-1}$s$^{-1}$) & 0.9 & 1.5 & 0.9 \\
$\alpha$ & 0.378 & 0.839 & 0.907\\
$\mathscr{E}_{c}/\mathscr{E}_{0} - 1$ (calculated) & 4.7\% & 21.6\% & 25.0\%\\
$\mathscr{E}_{c}/\mathscr{E}_{0} - 1$ (measured) & 4\%~\cite{Antonello:2020qht} & 16\%~\cite{Abratenko:2020bbx, Adams:2019qrr} & 19\% $\pm$ 4\%~\cite{Abi:2020mwi}\\
$1 - \mathscr{E}_{a}/\mathscr{E}_{0}$ (calculated) & 2.4\% & 11.9\% & 14.0\%\\
$1 - \mathscr{E}_{a}/\mathscr{E}_{0}$ (measured) & 2\%~\cite{Antonello:2020qht} & 8\%~\cite{Abratenko:2020bbx, Adams:2019qrr} & 11\% $\pm$ 2\%~\cite{Abi:2020mwi}\\
\bottomrule
\end{tabular}
\label{tab:sce}
\end{table}

Different schemes to measure and correct for the space charge effects were developed for the MicroBooNE~\cite{Abratenko:2020bbx, Adams:2019qrr} and ProtoDUNE~\cite{Abi:2020mwi} experiments. One first measures the spatial distortions using either the UV laser tracks~\cite{Adams:2019qrr} or the cosmic ray muon tracks~\cite{Abratenko:2020bbx, Abi:2020mwi}. Without the electric field distortion, the reconstructed laser track is straight in the liquid argon. By comparing the reconstructed track trajectory in the distorted electric field with the true laser trajectory, one can measure the spatial distortions. Even though the muon tracks are not straight because of multiple Coulomb scattering in the absence of space charge effects, one can still measure the average spatial distortion using many muon tracks. The most straightforward method is to use two tracks that nearly cross. The comparison of the true crossing point with the reconstructed crossing point gives the spatial distortion at that particular position. The crossing-track method was employed in Ref.~\cite{Abratenko:2020bbx}. It was not practical to use this method with MicroBooNE's laser system because the set of locations in the detector where two laser beams can nearly cross is limited due to the limited motion of the reflecting mirrors. Instead, the authors developed a iterative method to measure the spatial distortions using single laser tracks~\cite{Adams:2019qrr}. In the first ProtoDUNE SCE measurement, the spatial distortions were measured at the front, back, top and bottom surfaces of the TPC. The spatial distortions in the bulk of the TPC volume were obtained through interpolation of measurements at the surfaces and scaling of simulated spatial distortions~\cite{Abi:2020mwi}.

Once the spatial distortion map is determined throughout the TPC volume, the electric field distortions associated with SCE can be computed. First, the local electron drift velocity is calculated from the spatial distortion map~\cite{Abi:2020mwi}. Once the local drift velocity $v(x, y, z)$ is determined throughout the TPC volume, the electric field magnitude is obtained by using the relationship between the electric field and the drift velocity, which is a function of the liquid argon temperature. 

Both the spatial distortion map and the electric field map are used to correct the reconstructed track trajectory and the calorimetric measurement. The SCE affects the charge and energy loss per unit length ($dQ/dx$ and $dE/dx$) measurements in two ways. First, the spatial distortion changes the effective pitch $dx$ between two adjacent track hits. One can correct for this effect by recalculating the pitch $dx$ after correcting the two adjacent track hits using the spatial distortion map. Secondly, the distorted electric field changes the electron-ion recombination rate. One can correct for this effect by using the measured local electric field when converting $dQ/dx$ to $dE/dx$ using the recombination equation. More details on the recombination calibration will be discussed in Sec~\ref{sec:recomb}.

\subsection{Electron Attachment to Impurities}
In the presence of electronegative impurities, the concentration of free electrons $Q$ in liquid argon decreases exponentially with the drift time according to~\cite{Marchionni:2013tfa}
\begin{equation}
  \frac{dQ}{dt} = - k_{s}[S]Q,
\end{equation}
which leads to
\begin{equation}
  Q = Q_{0}e^{-k_{s}[S]t} = Q_{0}e^{-t/\tau_{e}},
\end{equation}
where $[S]$ is the concentration of electronegative impurities, $k_{s}$ is the electron attachment rate constant and $\tau_{e} = 1/(k_{s}[S])$ is the drift electron lifetime.

The liquid argon received from the supplier typically has contaminants of water, oxygen and nitrogen at the parts per million (ppm) level each. Water and oxygen capture drifting electrons and the concentration of these contaminants needs to be reduced by a factor of at least $10^{4}$ and maintained at this level to allow operation of the TPC. The purification system normally contains two filters~\cite{Benetti:1993mf, Curioni:2009rt, Adamowski:2014daa, Acciarri:2016smi, Abi:2020mwi}. The first filter that contains absorbent molecular sieve removes water contamination but can also remove small amounts of nitrogen and oxygen. The second filter that contains activated-copper-coated granules removes oxygen, and to a lesser extent, water. The nitrogen contaminant cannot be effectively filtered so the ultimate nitrogen concentration is set by the quality of the delivered argon. However, the nitrogen contaminant mainly suppresses the emission of scintillation photons and absorbs them as they propagate through the liquid argon~\cite{Jones:2013bca} and its impact on the drifting electrons is negligible. 

The attachment rate constants to oxygen have been measured as a function of the electric field strength~\cite{doi:10.1021/j100564a006}. For electric fields of less than a few 100 V/cm, the attachment constant is measured to be $k_{O_{2}} = 9\times10^{10}\ \textrm{M}^{-1}\textrm{s}^{-1} = 3.6\ \textrm{ppm}^{-1} \mu s^{-1}$, which corresponds to $\tau_{e}(\mu s)\approx300/[\textrm{O}_{2}] (ppb)$. For increasing electric fields, the attachment rate constant decreases, as observed in several other measurements~\cite{Swan:1963dla, Bettini:1990rj}. At an electric field of 1 kV/cm, $\tau_{e}(\mu s)\approx500/[\textrm{O}_{2}] (ppb)$.

No direct measurement of attachment rate for water in liquid argon is reported in literature. A direct relation between the water concentration in the vapor above the liquid argon and the electron drift lifetime in the liquid argon is observed: (Drift Lifetime)$\times$(Water Concentration) = Constant, as reported in Ref.\cite{Andrews:2009zza}. In the same measurement, the authors concluded the concentration of water in the liquid argon effectively limits the drift electron lifetime. 

\begin{figure}[htp!]
\centering
  \includegraphics[width=0.70\textwidth]{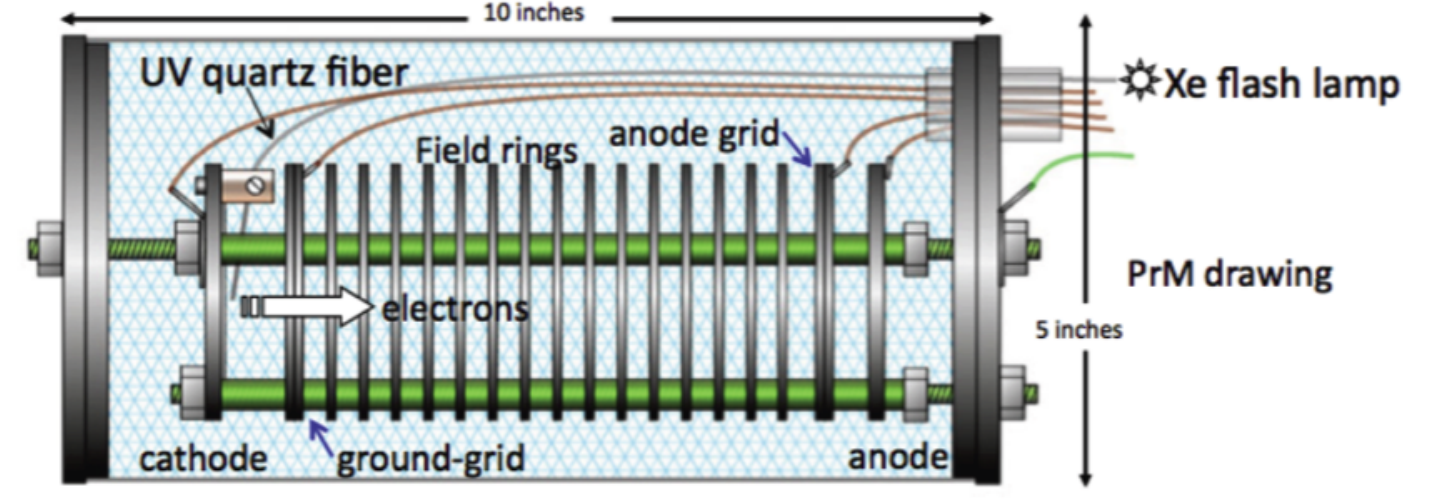}
\caption{
A drawing of a liquid argon purity monitor. This figure is from~\cite{Adamowski:2014daa}.
}
\label{fig:prm}
\end{figure}

The drift electron lifetime can be measured using the purity monitor~\cite{Carugno:1990kd, Amerio:2004ze, Amoruso:2004ti, Adamowski:2014daa, MicroBooNE:2016isc}. A purity monitor is a double gridded ion chamber immersed in the liquid argon volume, as shown in Figure~\ref{fig:prm}. The fraction of electrons generated via the photoelectric effect by the Xe lamp at the cathode that arrive at the anode $Q_{A}/Q_{C}$ after the electron drift time, $t$, is a measure of the electronegative impurity concentration and can be interpreted as the electron lifetime, $\tau_{e}$, such that
\begin{equation}
  Q_{A}/Q_{C} = e^{-t/\tau_{e}}.
\end{equation}
The SCE does not affect the operation of a purity monitor because the electrons are produced on the surface of the photocathode by the Xe lamp, and no ions are produced.

\begin{figure}[htp!]
\centering
  \includegraphics[width=0.49\textwidth]{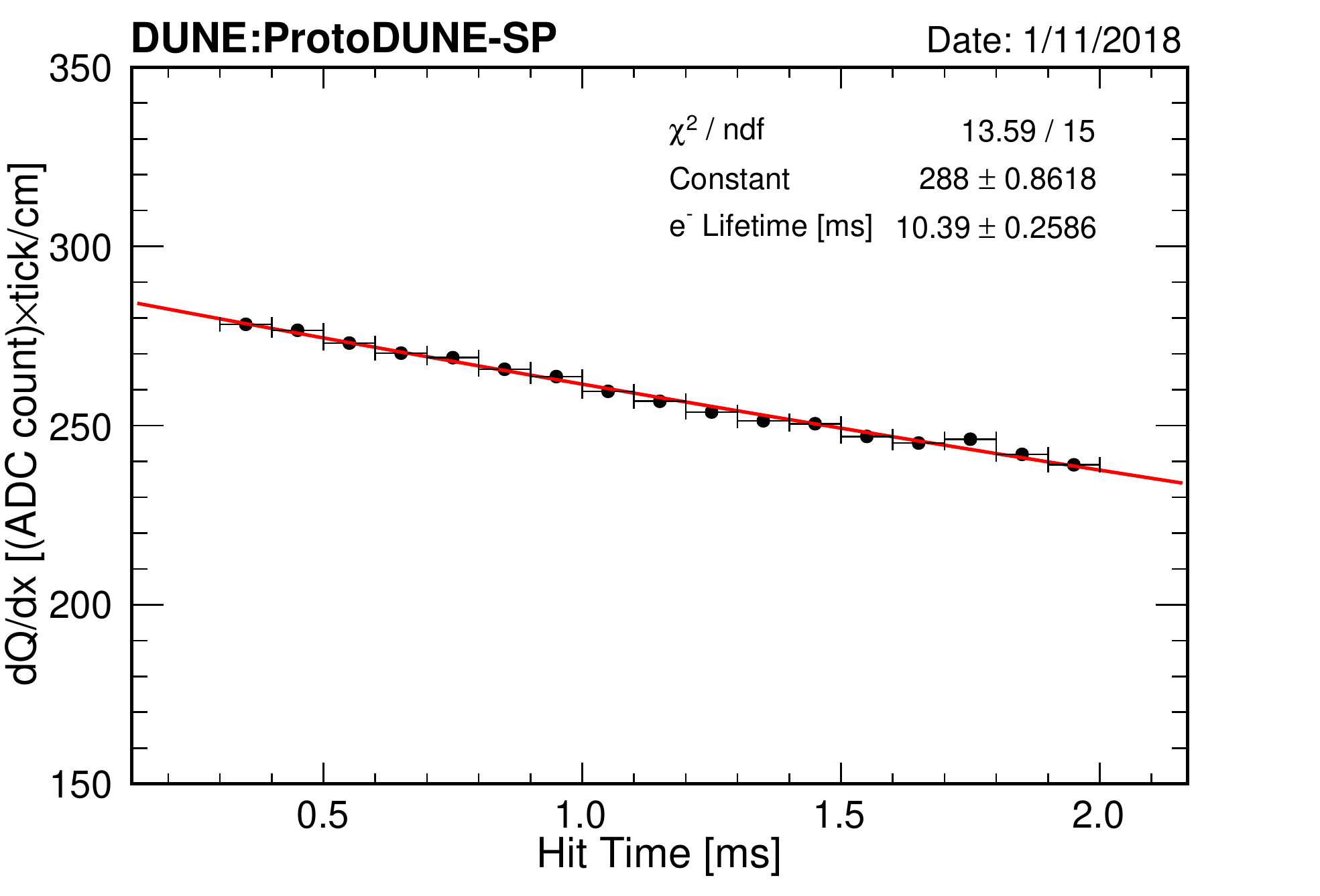}
  \includegraphics[width=0.49\textwidth]{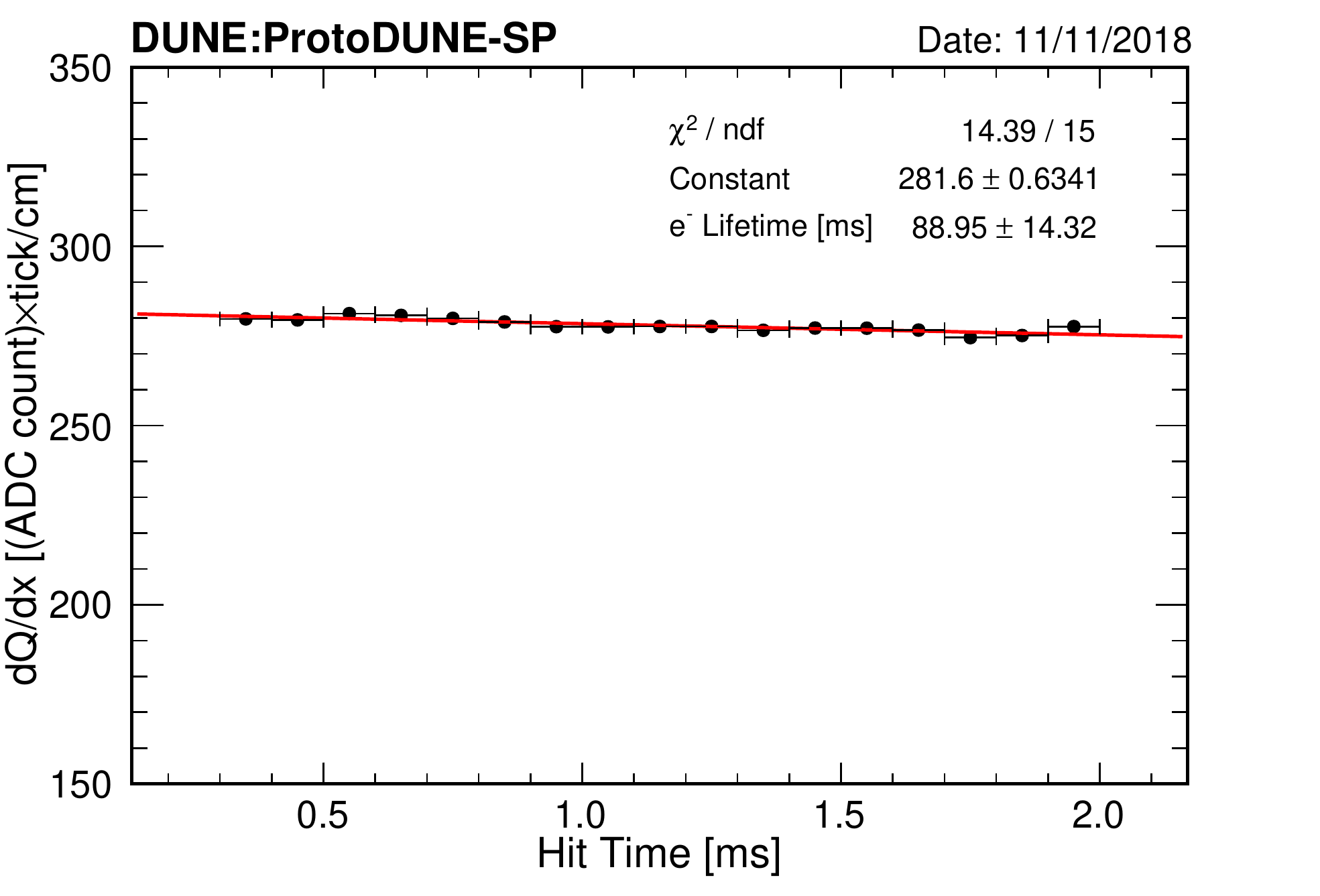}
\caption{
Plot the most probable value of the $dQ/dx$ distribution in ProtoDUNE as a function of the hit time, fit to an exponential decay function during a period of lower purity (left: $\tau_{e}$ = 10.4 ms) and during a period of higher purity (right: $\tau_{e}$ = 89 ms). This figure is from~\cite{Abi:2020mwi}.
}
\label{fig:lifetime}
\end{figure}

The drift electron lifetime can also be measured using cosmic ray muons recorded inside the TPC~\cite{Amoruso:2004ti, Bromberg:2015bqs, Anderson:2012vc, MicroBooNE:2017hgx, Acciarri:2019wgd, Abi:2020mwi}. The charge deposition per unit length measured by the muons is a function of the electron drift time for a constant drift electron lifetime:
\begin{equation}
  dQ/dx = (dQ/dx)_{0}e^{-t/\tau_{e}},
\end{equation}
where $(dQ/dx)_{0}$ is measured at the anode. For a LArTPC located on the surface, the SCE changes the measured $dQ/dx$ as discussed in Sec.~\ref{sec:sce}. Without correcting for the SCE, the measured electron lifetime is larger than the true electron lifetime because of the charge squeezing effect due to the spatial distortion and the change to the electron recombination due to the electric field distortion. ProtoDUNE developed a method to measure the drift electron lifetime using tracks tagged by the Cosmic-ray Tagging system (CRT). The selected TPC tracks are parallel to wire planes and perpendicular to the collection wire direction. Only the track segment in the central part of the TPC is used to minimize the spatial distortion caused by the SCE in the direction transverse to the nominal electric field. The change to the $dQ/dx$ measurement caused by the electric field distortion through recombination is small (less than 2\%) and corrected for using the measured electric field map. Figure~\ref{fig:lifetime} shows the lifetime measured in ProtoDUNE using CRT tagged muon tracks for two runs at different purity levels.

The two methods to measure the drift electron lifetime are complementary to each other. The purity monitors provide instant monitoring of the liquid argon purity but they measure purity in only a few locations outside the TPC. The muon method provides a direct measurement of the argon purity inside the TPC. The two methods may give different results of electron lifetime. The purity monitor normally operates at a much lower electric field ($\sim$ 25  V/cm) compared with the electric field in the TPC volume (500 V/cm). The rate constant for the attachment of electrons to the oxygen contaminant depends on the electric field, which means the measured lifetime using muons is higher than the one measured with the purity monitor if all contaminants are oxygen. The argon purity inside and outside the TPC can be different due to the fluid flow. The purity monitor measurement can be calibrated to provide the electron lifetime information inside the TPC, which is useful for the calibration of a LArTPC located deep underground (such as DUNE) where the cosmic flux is highly reduced. 

\subsection{Diffusion}

During the drift, the ionization electron cloud spreads, which is an effect known as diffusion. The diffusion of electrons in strong electric fields is generally not isotropic. In general, the diffusion in the direction of the drift field (longitudinal diffusion) is smaller than the diffusion in the direction transverse to the field (transverse diffusion). The longitudinal and transverse sizes of an electron cloud are given by:
\begin{equation}
  \sigma_{L}^{2}(t) = \left(\frac{2D_{L}}{v_{d}^{2}}\right)t,
\end{equation}
\begin{equation}
  \sigma_{T}^{2}(t) = \left(\frac{2D_{T}}{v_{d}^{2}}\right)t,
\end{equation}
where $D_{L}$ and $D_{T}$ are longitudinal and transverse diffusion coefficients, $t$ is the electron drift time, and $v_d$ is the electron drift velocity. The diffusion coefficients are given by the Einstein-Smoluchowski relation~\cite{https://doi.org/10.1002/andp.19053220806, https://doi.org/10.1002/andp.19063261405}:
\begin{equation}
  D = \frac{\varepsilon_{e}}{e}\mu_{e},
\end{equation}
where $\varepsilon_{e}$ is the mean electron energy, $e$ is the electron charge, and $\mu_{e}$ is the electron mobility, which is a function of the electric field. 
At relatively low electric field, the electrons gain so little energy from the field between the elastic atomic collisions that they come to thermal equilibrium with the medium. In this case, we have $\varepsilon_{e} = k\cdot T = 0.0075$ eV for argon at the nomal boiling point ($T$ = 87.3 K), where $k$ is the Boltzmann constant, and $\mu_{e} \approx 518$ cm$^{2}$/V/s~\cite{Li:2015rqa}. The corresponding coefficient is $D \sim 3.9$ cm$^{2}$/s, which is a lower limit to the diffusion coefficient. At strong electric fields the mean longitudinal electron energy increases while the electron mobility decreases. Therefore, the longitudinal diffusion coefficient is nearly a constant up to $\mathscr{E} = 1$ kV/cm. The ratio of the longitudinal to transverse diffusion coefficient is expressed as~\cite{Li:2015rqa}:
\begin{equation}
  \frac{D_{L}}{D_{T}} = 1 +\frac{\mathscr{E}}{\mu_{e}}\frac{\partial\mu_{e}}{\partial\mathscr{E}}.
\end{equation}
At low electric fields, the mobility is close to a constant, and we have $D_{L}/D_{T}\approx1$.

The longitudinal diffusion broadens the signal waveforms as a function of drift time while the transverse diffusion smears the signal between neighboring wires. A good understanding of the diffusion process is important because it can influence the event images and the accuracy of the drift coordinate measurement.

The ICARUS collaboration reported measurements of the longitudinal diffusion effect at four different electric fields: 100, 150, 250, 350 V/cm using a three-ton LArTPC with a maximum drift distance of 42 cm~\cite{Cennini:1994ha}. The longitudinal diffusion coefficient $D_{L}$ was derived from the analysis of the rise time of the signal on the collection wires. The resulting coefficients at different fields are consistent within the errors, which gives an average measurement of $D_{L} = 4.8\pm0.2$ cm$^{2}$/s. Li et al. from BNL reported measurements of $D_{L}$ between 100 and 2000 V/cm using a laser-pulsed gold photocathode with drift distance ranging from 5 to 60 mm~\cite{Li:2015rqa}. The ICARUS results show a good agreement with the prediction of Atrazhev and Timoshkin~\cite{689434}, while the BNL results are systematically higher than both.

Both MicroBooNE and ProtoDUNE are measuring the longitudinal diffusion coefficient $D_{L}$ by measuring the squared time width of a signal pulse, $\sigma_{t}^{2}$ as a function of drift time $t$:
\begin{equation}
  \sigma_{t}^{2}(t) \approx \sigma_{0}^{2} + \left(\frac{2D_{L}}{v_{d}^{2}}\right)t,
\end{equation}
where $\sigma_{0}$ is the signal width at anode, mostly determined by the electronics response discussed in Sec.~\ref{sec:electronics}. The results are expected to be published soon.

The diffusion process does not affect the charge reconstruction if hits are reconstructed with a variable time width and if the hit integral rather than the peak amplitude is used to measure the deposited charge since drift electrons are not lost by this process.

\subsection{Recombination}
\label{sec:recomb}
Electrons emitted by ionization are thermalized by interactions with the surrounding medium after which time they may recombine with nearby ions~\cite{Onsager:1938zz, https://doi.org/10.1002/andp.19133471205}. Electron-ion recombination introduces a non-linear relationship between $dE/dx$ and $dQ/dx$. The recombination effect can be measured using the stopping particles (muons or protons), which cover a wide range of $dE/dx$ values. For each point on a stopping track, $dE/dx$ is calculated from the distance to the track end (residual range)~\cite{Acciarri:2013met}, and $dQ/dx$ is calculated by converting the measured ADC counts to the number of electrons (discussed in ~\ref{sec:electronics}). The recombination effect can be parameterized by two empirical models. One is the Birks' model developed by the ICARUS collaboration~\cite{Amoruso:2004dy}:
\begin{equation}
\frac{dQ}{dx}(e/cm)=\frac{A_B}{W_{\textrm{ion}}}\left(\frac{\frac{dE}{dx}}{1+\frac{k_{B}}{\rho\mathscr{E}}\frac{dE}{dx}}\right),
\label{eqn:Birks_de_dx}
\end{equation}
where $A_{B}$ and $k_{B}$ are the two model parameters, $W_{\textrm{ion}} = 23.6$ eV is the ionization work function of arogn, $\mathscr{E}$ is the drift electric field, and $\rho$ is the liquid argon density. The other model is the modified Box model developed by the ArgoNeuT collaboration~\cite{Acciarri:2013met}:
\begin{equation}
\label{eqn:dq_dx_modbox}
\frac{dQ}{dx}(e/cm) =\frac{\ln(\frac{dE}{dx}\frac{\beta^{\prime}}{\rho\mathscr{E}}+\alpha)}{\frac{\beta^{\prime}}{\rho\mathscr{E}} W_{\textrm{ion}}},
\end{equation}
where $\alpha$ and $\beta^{\prime}$ are the two model parameters, and the other parameters are the same as in the Birks' model. In the detector calibration, $dE/dx$ is calculated by inverses of Eqs.~\ref{eqn:Birks_de_dx} and~\ref{eqn:dq_dx_modbox}:
\begin{equation}
  \frac{dE}{dx} = \frac{dQ/dx}{A_{B}/W_{\textrm{ion}}-k_{B}(dQ/dx)/(\rho\mathscr{E})},
  \label{eqn:de_dx_birks}
\end{equation}
or
\begin{equation}
  \frac{dE}{dx} = \frac{\rho\mathscr{E}}{\beta^{\prime}}(\exp(\beta^{\prime}W_{\textrm{ion}}(dQ/dx)/(\rho\mathscr{E}))-\alpha).
  \label{eqn:de_dx_box}
\end{equation}
The modified Box model describes data more accurately at higher $dQ/dx$. For $dQ/dx > (A_{B}\rho\mathscr{E})/(W_{\textrm{ion}}k_{B})$, $dE/dx < 0$ according to Eq.~\ref{eqn:de_dx_birks}. The LArIAT experiment combines both models for different ionization densities in the Michel electron measurement~\cite{Foreman:2019dzm}.

The recombination effects have been measured by the ICARUS~\cite{Amoruso:2004dy}, ArgoNeuT~\cite{Acciarri:2013met} and MicroBooNE~\cite{Adams:2019ssg} experiments at different electric fields. Table~\ref{tab:recomb} summarizes the recombination model parameters measured by the three experiments. We would like to point out that the two models with the best fit parameters are only valid in a given range of electric fields and $dE/dx$ values. Note the $k_{B}$ and $\beta^{\prime}$ parameters measured by the MicroBooNE experiment are lower than the ones measured by the other two experiments. Furthermore, the MicroBooNE recombination measurement was performed using an old version of simulation and before the SCE calibration was available. More details can be found in Ref.~\cite{Adams:2019ssg}.
\begin{table}[htp!]
\caption{The Birks and modified Box model parameters measured by the ICARUS, ArgoNeUT and MicroBooNE experiments}
\centering
\begin{tabular}{c|ccc}
\toprule
& \textbf{ICARUS}	& \textbf{ArgoNeuT} & \textbf{MicroBooNE}\\
\midrule
$\mathscr{E}$ (V/cm) & 200, 350, 500 & 481 & 273 \\
Sample & Stopping muons & Stopping protons & Stopping protons \\
Birks $A_{B}$ & $0.800 \pm 0.003$ & $0.806 \pm 0.010$ & $0.816 \pm 0.012$ \\
Birks $k_{B}$ ((kV/cm)(g/cm$^{2}$)/MeV) & $0.0486 \pm 0.0006$ & $0.052 \pm 0.001$ & $0.045 \pm 0.001$ \\
Box $\alpha$ & - & $0.93 \pm 0.02$ & $0.92 \pm 0.02$ \\
Box $\beta^{\prime}$ ((kV/cm)(g/cm$^{2}$)/MeV) & - & $0.212 \pm 0.002$ & $0.184 \pm 0.002$ \\
\bottomrule
\end{tabular}
\label{tab:recomb}
\end{table}

\subsection{Stopping Muon Calibration}
\begin{figure}[!ht]
\centering
\includegraphics[width=0.7\textwidth]{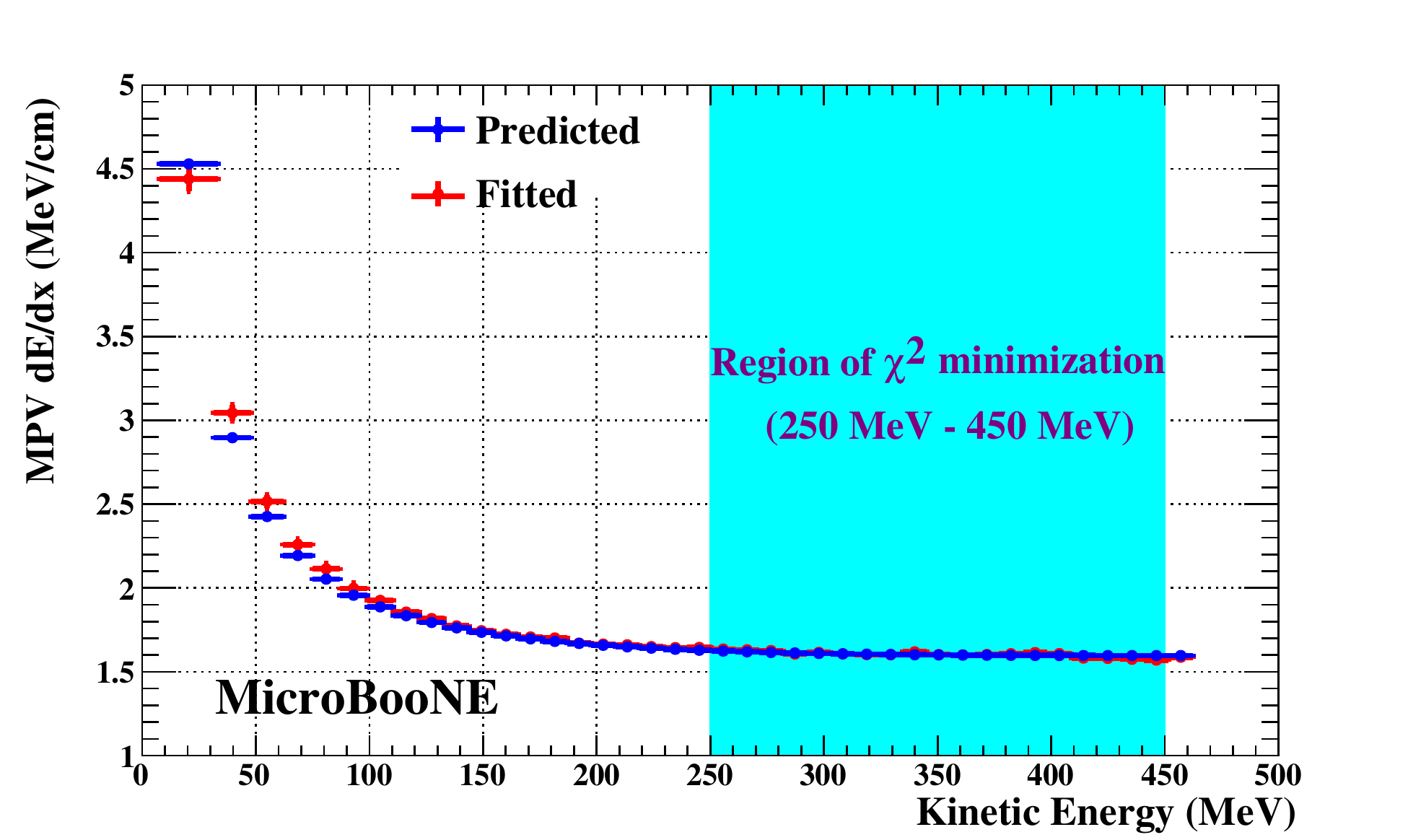} 
\caption{Comparison between the predicted and the measured most probable value $dE/dx$ for stopping muons in MicroBooNE data. This figure is from~\cite{Adams:2019ssg}.}
\label{fig:cali_plots} 
\end{figure}

There are generally two steps in converting the measured raw ADC counts to energy. The first step is to convert $\frac{dQ}{dx}$ (ADC/cm) to $\frac{dQ}{dx}$ ($e$/cm). This step employs the measured electronics gain and removes effects that cause non-uniformities in the detector response, including the SCE and the free electron attenuation. The second step converts $\frac{dQ}{dx}$ ($e$/cm) to $\frac{dE}{dx}$ (MeV/cm) using either the Birks or modified Box recombination model. However, there could be uncertainties in each step of calibration that affect the final calibrated $dE/dx$ quantity. One can introduce a calibration constant to correct the calibrated charge. This calibration constant is a scaling factor that applies to $\frac{dQ}{dx}$ ($e$/cm), which can be fine-tuned using the stopping muons. The kinetic energy of the muon at each track hit can be determined by the residual range. A portion of the track can be selected that corresponds closely to a minimum ionizing particle, for which $dE/dx$ is very well understood theoretically to better than 1\%. The calibration constant ensures the measured $dE/dx$ agrees with the prediction made by the Landau-Vavilov function~\cite{Bichsel:1988if} in the MIP region. Figure~\ref{fig:cali_plots} shows the measured $dE/dx$ values in MicroBooNE data after tuning the calibration constant, which show a good agreement with the prediction in the MIP region. This universal calibration constant can be used to calibrate the detector response to all particles. 

In principle, if the electronics gains are correctly measured by the pulser calibration system and all the effects that cause non-uniformities in the detector response are properly corrected for, the calibration constant derived using stopping muons should be exactly 1. The goal of the LArTPC calibration is to understand all the detector and physics effects in order to achieve an accurate and robust calorimetric measurement. 
\section{Conclusions}
In this article, we review the major effects that affect the calorimetric reconstruction in a LArTPC: electronics and field responses, space charge effects, electron attachment to impurities, diffusion and recombination. We also summarize the general methods to measure those effects and the results from different LArTPC experiments. Precise measurements of those effects improve the understanding of the detector response and energy resolution, which is crucial for reach physics sensitivity in the current and future LArTPC experiments. 

\vspace{6pt} 




\funding{Work supported by the Fermi National Accelerator Laboratory, managed and operated by Fermi Research Alliance, LLC under Contract No. DE-AC02-07CH11359 with the U.S. Department of Energy. The U.S. Government retains and the publisher, by accepting the article for publication, acknowledges that the U.S. Government retains a non-exclusive, paid-up, irrevocable, world-wide license to publish or reproduce the published form of this manuscript, or allow others to do so, for U.S. Government purposes.}

\acknowledgments{The author thanks Bruce Baller, David Caratelli, Flavio Cavanna, Tom Junk, and Stephen Pordes from Fermilab for their careful reading of the paper and for their helpful comments.}

\externalbibliography{yes}
\bibliography{citedb}



\end{document}